\documentclass[twocolumn,
superscriptaddress,
amsmath,amssymb,aps,showkeys,showpacs,
twoside,final,secnumarabic,
nofootinbib]{revtex4-2}

\usepackage[paperwidth=205mm,paperheight=290mm,top=17mm,bottom=25mm,
inner=17mm,outer=17mm,
twoside]{geometry}

\usepackage{cmap} 
\usepackage[T1,T2A]{fontenc}
\usepackage[utf8]{inputenc}
\usepackage[russian,english]{babel}
\usepackage{color}
\usepackage{graphicx}
\usepackage{dcolumn}
\usepackage{bm} 
\usepackage[unicode=true,colorlinks=true,linkcolor=magenta, urlcolor=blue, citecolor = blue,breaklinks]{hyperref}
\usepackage{multirow}
\usepackage{url}
\usepackage{breakurl}
\DeclareGraphicsExtensions{.eps}

\newcount\issue
\newcount\Vol
\newcount\numb
\usepackage{fancyhdr} 
\def\Vol{\textbf{80}}
\def\numb{x}
\begin{document}
\title{Modern Stellar Astronomy \\[20pt]
Search for galactic structures and study of their kinematics \\in the Gaia era}

\def\addressa{Cental (Pulkovo) Astronomical Observatory RAS, 65 bld.1, Pulkovskoe shosse, Sankt-Peterburg 196140, Russia}
\def\addressb{address 2}

\author{\firstname{V.~V.}~\surname{Bobylev}}
\email[E-mail: ]{bob-v-vzz@rambler.ru}
\affiliation{\addressa}
 \author{\firstname{A.~T.}~\surname{Bajkova}}
\affiliation{\addressa}
\email[E-mail: ]{anisabajkova@mail.ru}

\received{xx.xx.2025}
\revised{xx.xx.2025}
\accepted{xx.xx.2025}

\begin{abstract}
The characteristics of Gaia catalogues, such as trigonometric parallaxes and proper motions of stars, are discussed. Radial velocities of stars are also important for studying spatial motions. The most important results of the kinematics analysis of stellar groups from the nearest vicinity of the Sun are noted, where Gaia data allow estimating a number of parameters with unprecedentedly high accuracy. The issues related to the rotation of the Galaxy and its spiral structure are touched upon. The properties of the Radcliffe wave are discussed in the light of new data. An answer is obtained to the question of the existence of an analogue of the Radcliffe wave in other places of the Galaxy. The discovery of a phase spiral in the $z-V_z$ plane, made using Gaia data, is noted.
\end{abstract}

\pacs{Suggested PACS}\par
\keywords{Stellar kinematics, Galaxy rotation, spiral structure, Radcliffe wave, phase spiral \\[5pt]}

\maketitle\thispagestyle{fancy}

\section{Introduction}\label{intro}
A number of interesting results have been obtained in the last decade that are important for understanding the structure and kinematics of the Galaxy. This is largely due to the advent of mass astrometric and photometric measurements obtained as a result of the Gaia space experiment \cite{1},\cite{2}. By now, the trigonometric parallaxes and proper motions have been determined for approximately one and a half billion stars. For bright stars $7^m<G<14^m$, the parallax errors average less than 25 microarcseconds, $\mu$as), and the errors in each component of the proper motion are less than 25 microarcseconds per year ($\mu$as/yr). For fainter stars, down to $G\sim21^m$, the errors in all three quantities increase significantly. It is expected that in the final version of the catalog, random measurement errors will be reduced by two to three times. The radial velocities are important for a full-fledged analysis of stellar kinematics. To date, such velocities have been measured for approximately 30 million stars in the Gaia catalogue.

\section{\label{sect-1}The closest stellar groups to the Sun}
\begin{figure*}[t]
\includegraphics[width=0.80\textwidth]{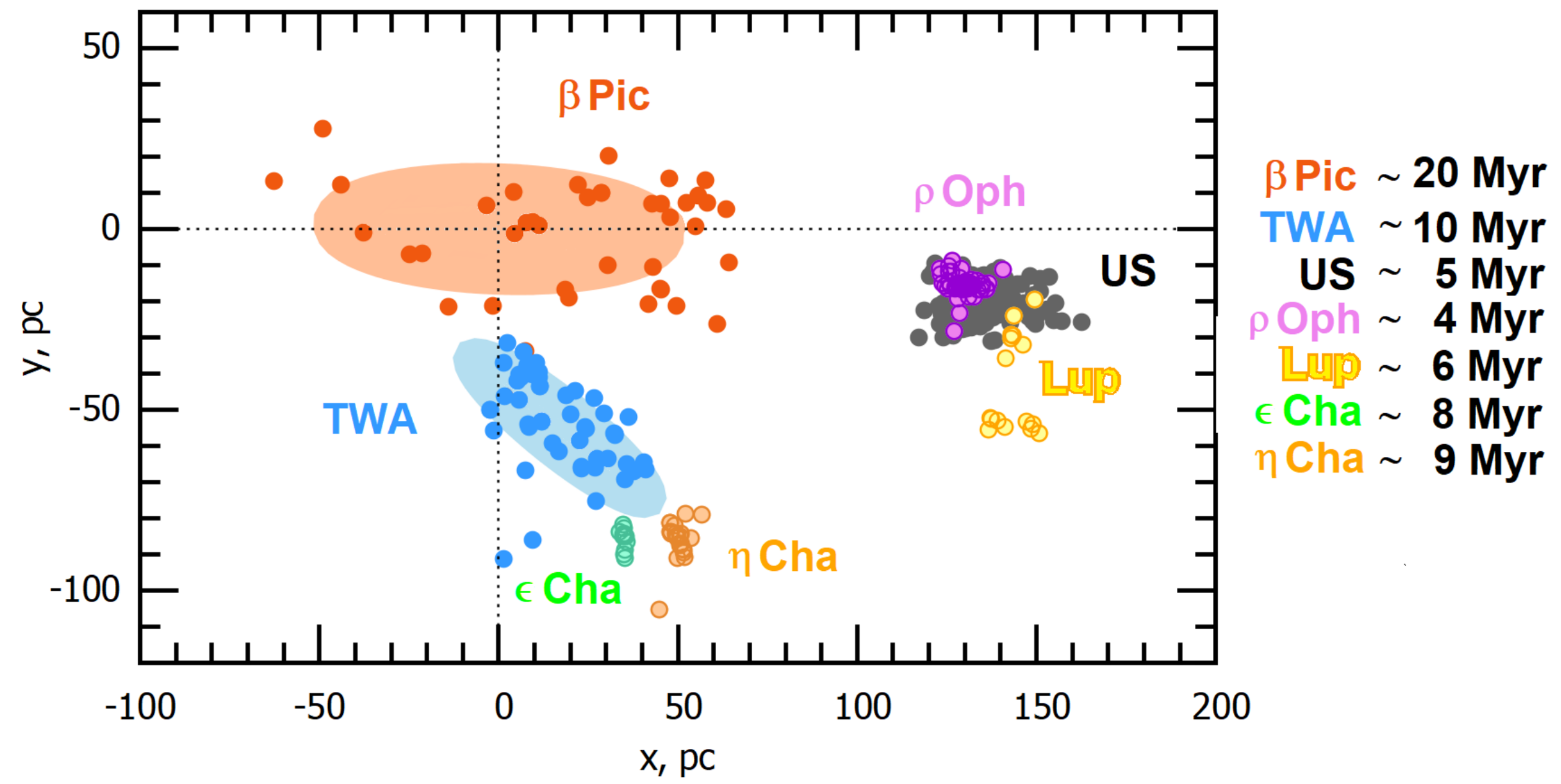}
\caption{\label{fig:XY} A number of young stellar associations and moving clusters closest to the Sun.}
\end{figure*}

The distribution of some young stellar associations and moving clusters closest to the Sun in projection onto the galactic plane $xy$ is given in Fig.~\ref{fig:XY}. Each point in the figure represents a real star from the Gaia~DR3 catalog with measured parallax and proper motion. For each star, the radial velocity value is also known, taken from ground-based or space measurements.

Let us note the TW~Hya association. A detailed analysis of the kinematics of this association was performed in \cite{3} and \cite{4}. Kinematic estimates of the age, $t$, of this association were obtained in two ways. The first method involved an analysis of stellar trajectories integrated backwards in time, which yielded an estimate of $t=4.9\pm1.2$~Myr. The second method showed from an analysis of the instantaneous velocities of the stars that there is a volumetric expansion of the stellar system with an angular velocity coefficient of $K_{xyz}=103\pm12$~km/s/kpc. Based on this effect, the time interval from the beginning of the expansion of the TW~Hya association to the present moment was found to be $t=9.5\pm1.1$~Myr.

For many stellar associations and clusters, the effects of expansion or compression, occurring either along one direction or occurring in one plane (usually in the $xy$ plane), have long been known. The volume expansion of the TW~Hya association was first found in \cite{3} and confirmed in \verb+\cite{4}+.

An analysis of the kinematics of numerous small stellar groups in the US association, performed in a recent paper \cite{5} led the authors to the necessity of changing the model of star formation within the US association. If earlier the model of sequential star formation was accepted for this association, now it is necessary to accept the model of stochastic star formation for it.

\section{\label{sect-2}Radcliffe Wave}
Based on the catalog of molecular clouds \cite{6} with highly accurate estimates of distances to them (with an average error of about 5\%), the Radcliffe wave was discovered in \cite{7}. This structure is a narrow chain of clouds stretched in a line with a length of $\sim$2.7~kpc in the galactic plane $xy$. This chain is inclined to the $y$ axis by an angle of $30^\circ$. Its main feature is a clearly visible wave-like nature of the distribution of clouds in the vertical direction. The maximum value of the coordinate $z\sim160$~pc is observed in the immediate vicinity of the Sun.

The wave-like behavior of vertical coordinates is confirmed in the distribution of interstellar dust, molecular clouds, masers and radio stars, T-Tauri stars, massive OB stars, and young open star clusters.

In paper \cite{8}, based on spectral analysis of masers with trigonometric parallaxes measured by the VLBI method, it was perhaps shown for the first time that in the Radcliffe wave, in addition to vertical coordinate perturbations, vertical velocity perturbations with an amplitude of $5.1\pm0.7$~km/s are observed. In paper \cite{9}, the presence of vertical velocity perturbations with a magnitude of about $5$~km/s was confirmed from an analysis of a large sample of young open star clusters. These authors also discovered wave motion toward the galactic anticenter, as well as a small tangential shift of the wave toward the rotation of the Galaxy.

The origin of the Radcliffe wave is still unclear. For example, in the work \cite{10} the origin of this wave is associated with the Kelvin--Helmholtz instability, which occurs in the galactic disk due to the difference in rotation speeds of the dark matter halo and the disk. Various authors discuss the hypothesis of the impact of an external impactor on the galactic disk, which could be a dwarf satellite galaxy of the Milky Way, a massive clot of dark matter, or a globular cluster. According to a number of authors \cite{9},\cite{11} the Radcliffe wave could have arisen as a result of the impact of shock waves from several supernovae and their stellar winds during the formation of the Local Bubble or the North Polar Spur. At the same time, the most important question of what role the magnetic field plays in the formation of bubbles is still open.

\begin{figure*}[t]
\includegraphics[width=0.8\textwidth]{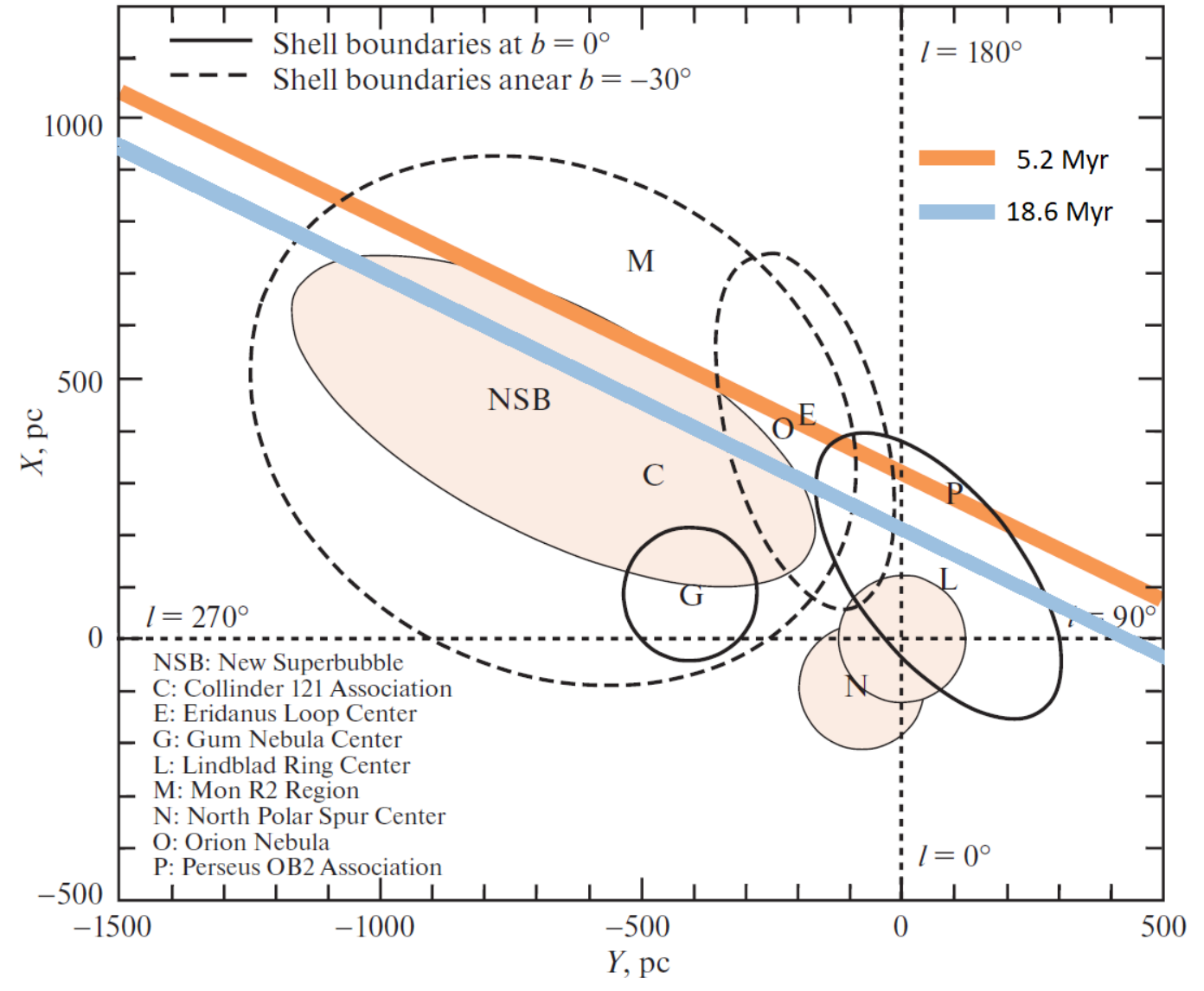}%
\caption{\label{fig:bubbles} Distribution of known bubbles (circles) inflated by supernova explosions on the galactic xy plane, as well as the positions of two samples of open star clusters (orange and blue lines) of different ages.}
\end{figure*}

Fig.~\ref{fig:bubbles} shows the distribution of known bubbles inflated by supernova explosions projected onto the galactic plane $xy$, as well as the positions of two samples of open star clusters of different ages. As can be seen from this figure, the bubble centers are grouped on average closer to the positions of older open star clusters. That is, along the distribution line of open star clusters in which the most massive OB stars have already evolved and exploded. This figure shows that the appearance of bubbles is a consequence of the development and radial (towards the galactic anticenter) motion of the Radcliffe wave, and not the cause of its occurrence due to the impact of shock waves on the initial matter from which it was formed.

\section{\label{sect-3}Radcliffe Wave and the Spiral Structure of the Galaxy}
In the paper \cite{12} a narrow chain of masers 3--4 kpc long, stretching in the direction of about $40^\circ$, passing from a segment of the Carina-Sagittarius spiral arm to the Scutum arm was studied in order to test the hypothesis \cite{13} that this chain may be an analogue of the Radcliffe wave. However, no significant periodic perturbations of vertical coordinates and velocities characteristic of the Radcliffe wave were found in this structure. That is, the chain of masers between segments of the Carina and Scutum spiral arms is not an analogue of the Radcliffe wave.

\begin{figure*}[t]
\includegraphics[width=1.0\textwidth]{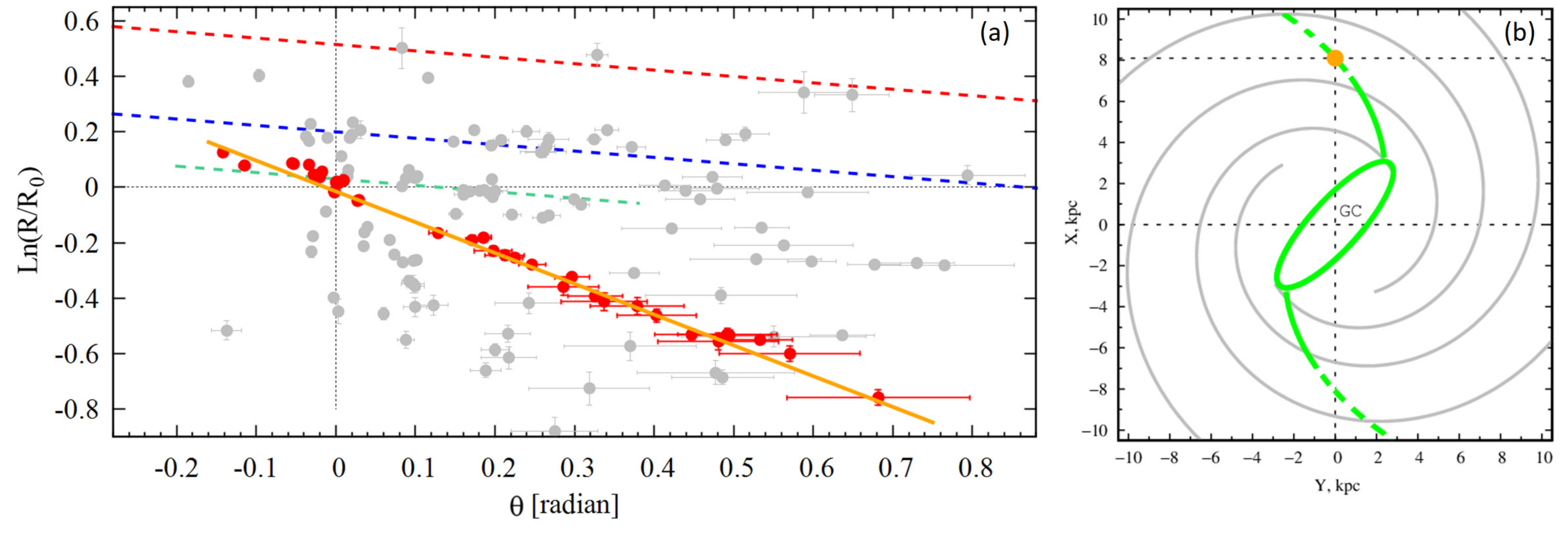}
\caption{\label{fig:masers} Masers with measured trigonometric parallaxes in projection onto the plane $\theta-\log (R/R_0)$~(a) and our model with two segments of spiral arms extending from the ends of the bar with twist angles of $-48^\circ$~(b).}
\end{figure*}

However, as a byproduct in the work \cite{12} it was shown that a narrow chain of masers extended from the Sun to the Scutum arm and may indicate the presence of a segment of a spiral arm with a twist angle of $-48^\circ$. Fig.~\ref{fig:masers} shows masers with measured trigonometric parallaxes in projection onto the $\theta-\log (R/R_0)$ plane, as well as the proposed model with two segments of spiral arms with twist angles of $-48^\circ$ extending from the ends of the bar.

In the model proposed in the work \cite{12} (see Fig.~\ref{fig:masers}~(b)) it is important that the segments of the spiral arms with a twist angle of $-48^\circ$ are part of the galactic bar, which rotates rigidly with an angular velocity of 40--50 km/s/kpc. In this case, for example, in the region of the Sun the difference in linear rotation velocities will reach about 100~km/s, which can have a great significance for the processes of star formation, for the shape of the Carina-Sagittarius spiral arm.

\section{\label{sect-4}Phase Spiral}
According to the Gaia DR2 catalog, a phase spiral was discovered in the $z-V_z$ plane in \cite{14}. Such a structure of this plane had never been studied before. According to the Gaia DR3 data, in \cite{15} this picture was clarified. These authors came to the conclusion that the Galaxy disk was disturbed 300--900 million years ago, which they associate with the previous pericentric passage of the Sagittarius dwarf galaxy.

As the authors of the phase spiral discovery write, “These results mark a new era in which, by modeling the richness of phase space substructures, we can determine the structure of the Galaxy and the characteristics of the perturbers that have had the greatest impact on our home in the Universe.”

\begin{figure*}[t]
\includegraphics[width=0.95\textwidth]{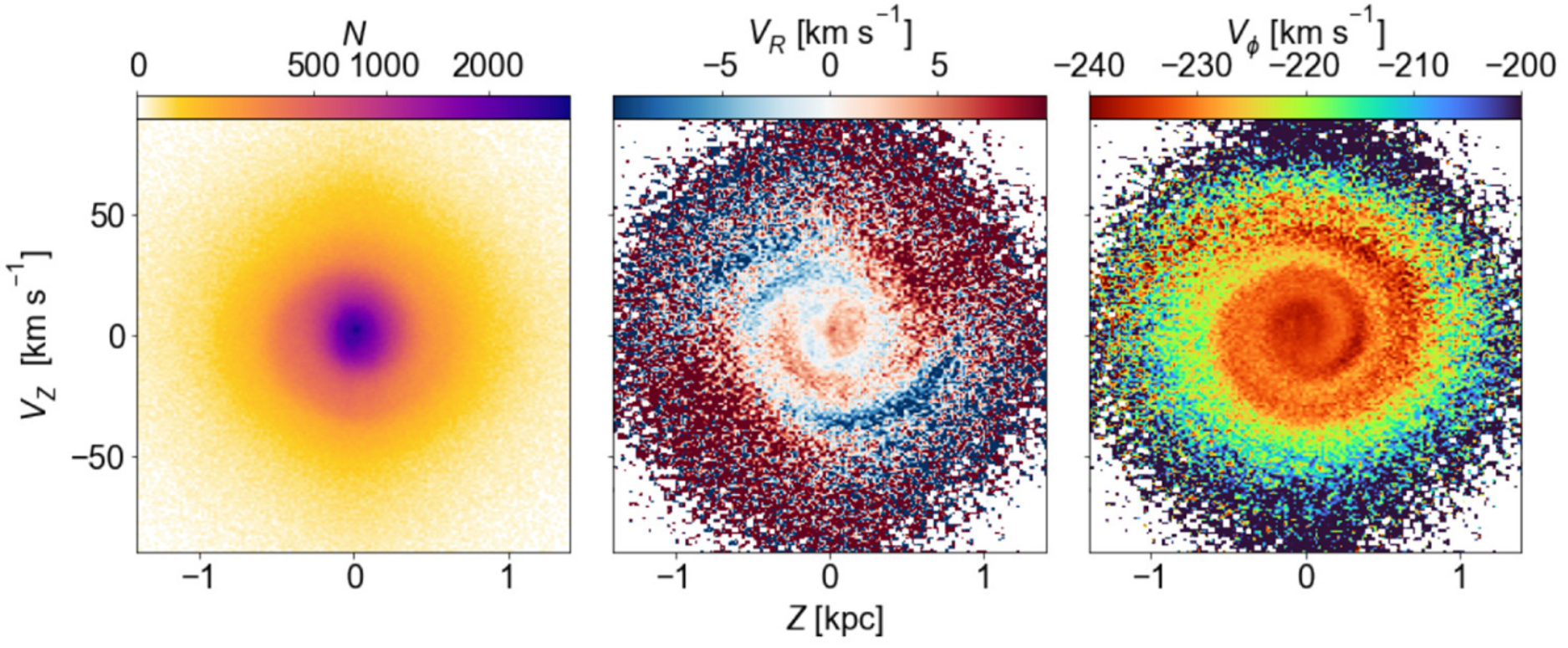}
\caption{\label{fig:Phase-spiral} Velocities $V_z$ as a function of coordinate $z$. Figure taken from Antoja, et al. (2023).}
\end{figure*}

Fig.~\ref{fig:Phase-spiral} shows a Phase spiral in the solar neighborhood with Gaia DR3 data. A large number of studies have been devoted to the study and modeling of this phenomenon. However, there is no complete understanding of the origin of the phase spiral yet.

\section{Conclusion}\label{Concl}
Some of the most interesting results, from our point of view, obtained using the Gaia space experiment data, devoted to the study of the structure and kinematics of the Galaxy are described. In particular, the results of the kinematic analysis of the young (younger than 20 million years) stellar groups closest to the Sun, for which the use of Gaia data is most effective, are noted. They are located no further than 200 pc from the Sun. Of great interest to researchers is the recently discovered Radcliffe wave near the Sun, the nature of which is currently unclear. The Radcliffe wave length is about 2.5 kpc, it is part of the so-called Local system (Orion arms). Its main feature is the presence of wave-like oscillations of the vertical coordinates $z$ and vertical velocities $V_z$. Some issues related to the spiral structure of the Galaxy are touched upon. Finally, according to the Gaia data, a phase spiral was revealed in the $z-V_z$ plane, indicating a large-scale instability of the galactic disk. The emergence of such a spiral is associated with gravitational disturbances caused by the absorption of a dwarf satellite galaxy of the Milky Way by the Galaxy in the past.

%
\section*{FUNDING}
The study was conducted under the state assignment of the Central (Pulkovo) Astronomical Observatory of Russian Academy of Science.


\section*{CONFLICT OF INTEREST}
The authors of this work declare that they have no conflicts of interest.

\end{document}